\documentclass[10pt,twocolumn,prX,showpacs,preprintnumbers,amsmath,amssymb,floatfix]{revtex4}
\usepackage{graphicx}
\usepackage{dcolumn}
\usepackage{bm}
\begin{document}

\preprint{APS/123-QED}

\title{A comparative study of the radius of sensitivity of the  
optical model\\ potentials for $^{6}$Li+$^{58,64}$Ni
and $^{16}$O+$^{58,64}$Ni}

\author{Mili Biswas}
\email{mili.biswas@saha.ac.in}
\affiliation{
Saha Institute of Nuclear Physics, 1/AF, Bidhan Nagar,
Kolkata-700 064, INDIA
}

\date{\today}

\begin{abstract}
 Radii of sensitivity were estimated for the $^{6}$Li+$^{58,64}$Ni system
at energies near the Coulomb barrier. For comparison purposes, such radii were also
estimated for stable $^{16}$O scattered from same target isotopes. The elastic
scattering data were analysed with folded real potential generated from DDM3Y
nucleon-nucleon interaction and an imaginary potential of volume Woods-Saxon form.
The most sensitive radii for $^{16}$O+$^{58,64}$Ni system are found to be 
energy independent and close to the strong absorption radius. For $^{6}$Li  
projectile, unlike its strongly bound counterpart, the crossing radius 
increases with decreasing energy. However, no two crossing situation
has been observed for both $^{6}$Li+$^{58,64}$Ni and $^{16}$O+$^{58,64}$Ni systems at the top of the barrier.

\end{abstract}

\pacs{25.60.Bx, 24.10.Ht, 27.20.+n}

\maketitle

 The threshold anomaly is a well known phenomenon
observed in case of heavy-ion scattering systems \cite{sat1} at low bombarding energies.
It refers to a strong variation of the real interaction potential with 
incident energies close to the Coulomb barrier. It is connected with the
increasing strength of the imaginary potential corresponding to the increasing 
availability of local energy to excite reaction channels in the same energy 
domain. The connection is through a dispersion relation [2,3] 
that arises from the causality in heavy ion collision. The dispersion integral
involves the real and imaginary components that need to be evaluated at a certain
radius value while investigating the energy dependence of the polarization potentials.
The general convention is to evaluate these quantities at the strong 
absorption radius. However, the question is whether the so-called strong 
absorption radius corroborates with the most sensitive radius or not as the
bombarding energy decreases? 
Roubos, et al., \cite{rou} have recently studied the scattering of $^6$Li and $^{16}$O
from heavy mass target $^{208}$Pb to investigate the radius of sensitivity 
for these systems.
The authors observed that for the tightly bound systems the appropriate radius of evaluation of dispersion relation is the strong absorption radius but for weakly bound systems that is not the case.
It is therefore important to ascertain the radial region of sensitivity of the
potentials before making use of the dispersion relation. Work has also been carried out
in this direction in Refs.[5-8]. 
In this context we present a systematic
study of the elastic data of $^{6}$Li and $^{16}$O projectiles on two different 
isotopes of medium mass Ni target to determine the radial region of the potential sensitivity
and to identify the difference in observation for the weakly and strongly bound nature of the
projectile as the target mass decreases.

 The elastic angular distributions of the system $^{6}$Li+$^{64}$Ni 
have been measured in an experiment performed at TIFR, Mumbai using the BARC-TIFR 
Pelletron Facility over the energy range 13-26MeV \cite{mb}. We have reanalysed 
the existing data of the system $^{6}$Li+$^{58}$Ni \cite{pf}.  
For $^{16}$O+$^{58,64}$Ni system we have used the data 
of Ref.[5,11]. In Fig.1 and Fig.2, the elastic angular distributions at some of the energies have been shown. 

\begin{figure}
\vskip -1.3cm
\resizebox{\columnwidth}{!}{\includegraphics{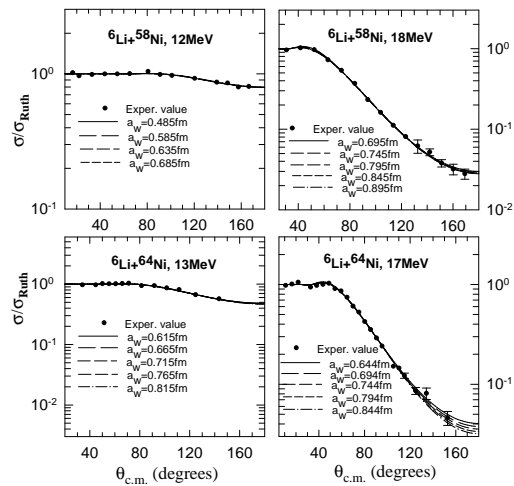}}
\vskip -3.0cm
\caption{\label{fig1}Elastic scattering angular distributions for the system
$^{6}$Li+$^{58,64}$Ni}
\end{figure}
 
\begin{figure}
\vskip -1.0cm
\resizebox{\columnwidth}{!}{\includegraphics{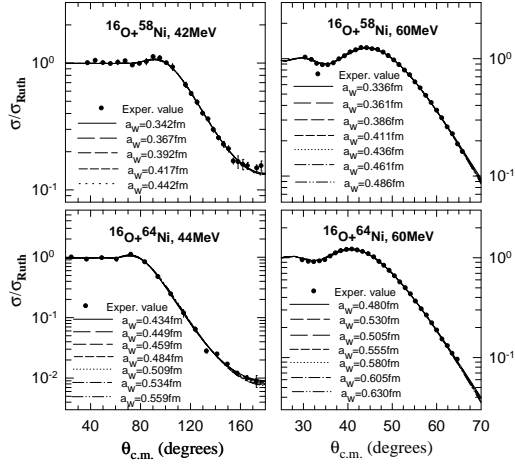}}
\vskip -3.0cm
\caption{\label{fig2}Elastic scattering angular distributions for the system
$^{16}$O+$^{58,64}$Ni}
\end{figure}

To investigate the radius of potential sensitivity and its possible variation
with incident energy, we have analyzed the 12,14,16,18 and 20MeV data of $^{6}$Li+$^{58}$Ni,
and 13,14,17,19 and 26MeV data of $^{6}$Li+$^{64}$Ni. The plots 
 for crossing point radius at 14,19 and 26MeV for 
$^{6}$Li+$^{64}$Ni are shown in Fig.3. 

\begin{figure}
\vskip -1.5cm
\includegraphics[scale=0.5]{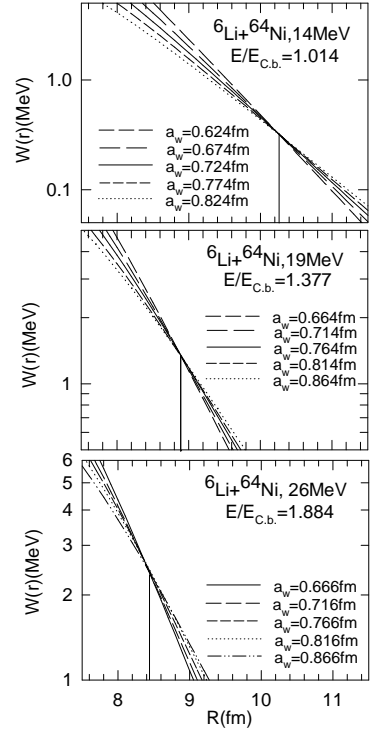}
\vskip -2.0cm
\caption{\label{fig3}Crossing radii for the system
$^{6}$Li+$^{64}$Ni at different energies. Coulomb barrier is 13.8MeV in lab frame according to Broglia and Winther \cite{bw}}
\end{figure}

\begin{figure}
\vskip -1.5cm
\includegraphics[height=12cm,width=8.6cm,angle=0]{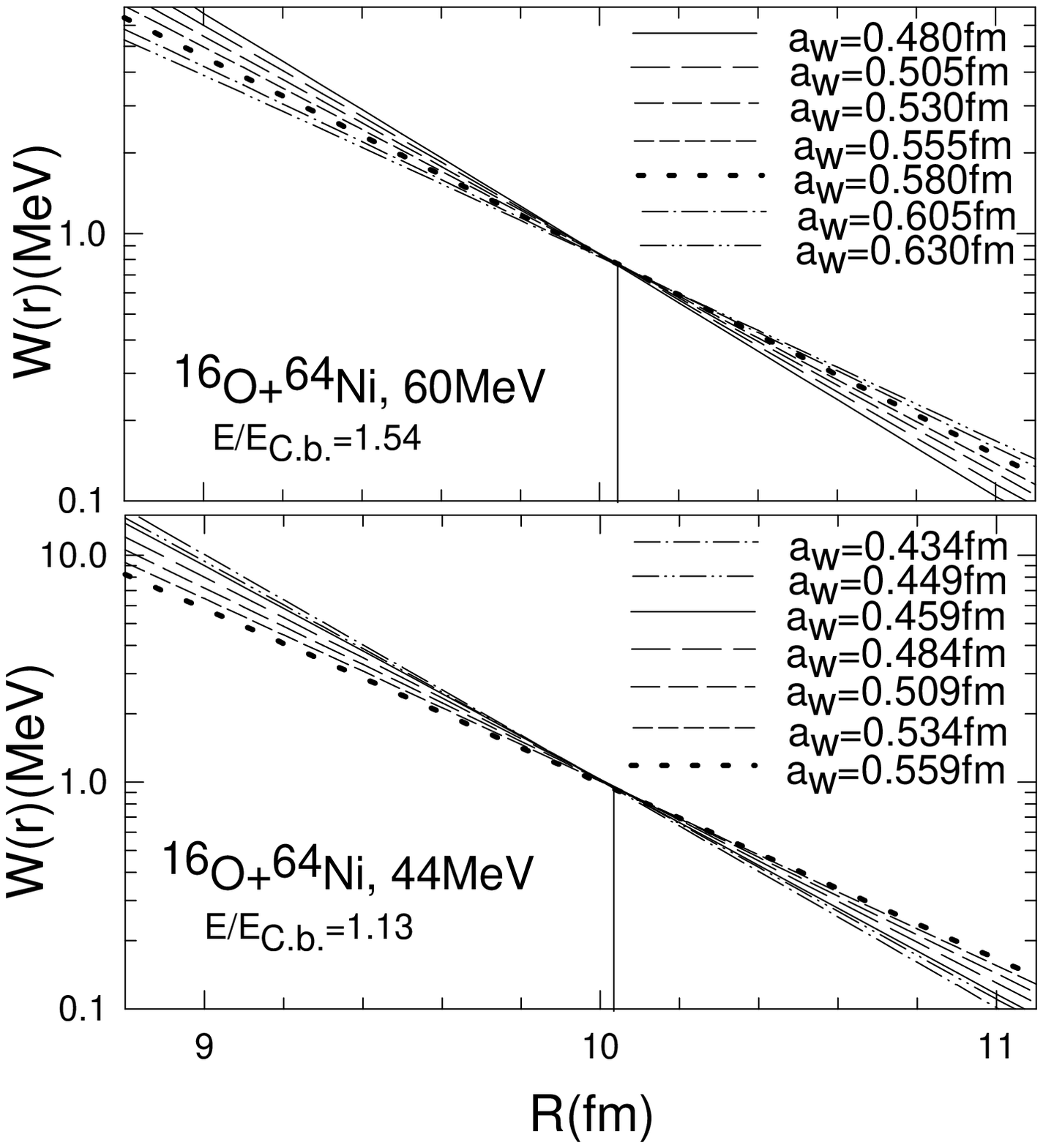}
\vskip -3.0cm
\caption{\label{fig4}Crossing radii for the system
$^{16}$O+$^{64}$Ni, $E_{C.b.}$ in lab=38.85MeV \cite{bw}}
\end{figure}

For comparison purpose the same plots at 44 and 60MeV for $^{16}$O+$^{64}$Ni have been shown in Fig.4.
All the new and existing elastic scattering data were analysed consistently
in terms of the optical model potential. The model potential $U_{mod}$(r) in the present
study has the form
\begin{equation}
U_{mod}(r)={\lambda_r}V_{fold}(r)+iW_v(W_0,r_w,a_w;r).
\end{equation}
$V_{fold}$(r) is the double folded potential and $W_v$ is the imaginary volume
Woods-Saxon potential. The renormalization factor $\lambda_r$ simulates the 
effect of $\Delta$V, the real part of the polarization potential 
related to the imaginary component as
\begin{equation}
\Delta {V(r;E)}= \frac{P}{\pi} \int \frac{W(r;E')}
{E'-E}dE'
\end{equation}
where P denotes the principal value.

The double-folded potentials
were calculated with the nickel mass densities obtained from Ref.\cite{rip}
and $^6$Li density by unfolding the parametrized charge density from 
Ref.\cite{sat2}.
The neutron density of $^6$Li was assumed to have the same shape as the 
proton density. Density of $^{16}$O was again taken from Ref.\cite{rip}. 
The M3Y nucleon-nucleon interaction in DDM3Y [14,15]
convention was used for the calculation that includes an intrinsic energy dependence
through a multiplicative factor of g(E)=(1-0.002E).

To obtain the best fit parameters of the potentials analysis was started with the
highest energy data for all the systems. At the highest energy with the model potential $U_{mod}$(r) we performed an initial search over all the
four parameters ($\lambda_r$, $W_0$, $r_w$, $a_w$) simultaneously. Subsequently, the imaginary radius parameter obtained from the initial search was
kept fixed. The best fit, determined by $\chi^2$ minimization, was found by searching over the real renormalization
factor and the imaginary strength while gridding over the imaginary diffuseness $a_w$. Same search procedure was adopted
for all the incident energies. The radius parameter was held fixed throughout assuming that the change in the value of
$r_w$ due to change in incident energy is not so significant. The range of the diffusivity parameter was determined 
by the condition of similar $\chi^2$/N. For the systems $^{6}$Li+$^{58,64}$Ni 
we have considered sets of potential parameters generating equally good fit to the elastic scattering angular distributions shown in Fig.1 with different diffusivities.

It was observed that if we varied the diffusivities beyond the range of values shown, the different sets of potential parameters would not intersect proving that they are no good potentials describing the elastic scattering angular distributions properly. 
The same procedure has been performed for the system $^{16}$O+$^{58,64}$Ni. In Fig.2, all the calculated angular distributions with different diffusivities have been shown. Though the $\chi^2$/N values of the fits vary within the range of 2$\chi^2_{min}$/N, the corresponding fits are quite good. The observed departure at large angles are well within the error limit of the data. For these systems the crossing points are close to the strong absorption radius where the various reactions are expected to take place.   
The search code ECIS94 \cite{raynal} was used to perform the 
model calculations. The optical model potential parameters 
obtained following the above search procedure along with the $\chi^2$/N (N
denotes the number of data points) values and the reaction cross sections $\sigma_{R}$ 
with different diffusivities have been given in TablesI-II.

\begin{table}
\caption{\label{tab:Table1}Potential parameters for $^6$Li+$^{64}$Ni}
                                                                                                                             
\begin{tabular}{ccccccc}\hline
                                                                                                                             
E(MeV) & $N_R$ & WS & $R_w$ & $a_w$ & $\chi^2$/point & $\sigma_R$(mb)\\
\hline
14.0 & 0.97 & 87.90 & 6.753 & 0.624 & 4.161 & 362.73\\
     & 0.85 & 58.52 & 6.753 & 0.674 & 4.065 & 365.54\\
     & 0.75 & 41.13 & 6.753 & 0.724 & 4.036 & 369.59\\
     & 0.67 & 29.95 & 6.753 & 0.774 & 4.063 & 374.60\\
     & 0.60 & 22.71 & 6.753 & 0.824 & 4.155 & 379.22\\
19.0 & 0.65 & 34.93 & 6.753 & 0.664 & 3.236 & 913.89\\
     & 0.58 & 28.15 & 6.753 & 0.714 & 3.107 & 932.22\\
     & 0.51 & 23.56 & 6.753 & 0.764 & 3.091 & 953.11\\
     & 0.44 & 19.81 & 6.753 & 0.814 & 3.192 & 972.34\\
     & 0.38 & 17.29 & 6.753 & 0.864 & 3.431 & 997.94\\
26.0 & 0.72 & 32.58 & 6.753 & 0.666 & 0.633 & 1401.56\\
     & 0.67 & 27.70 & 6.753 & 0.716 & 0.574 & 1431.90\\
     & 0.60 & 24.46 & 6.753 & 0.766 & 0.536 & 1467.47\\
     & 0.53 & 21.87 & 6.753 & 0.816 & 0.522 & 1504.00\\
     & 0.45 & 19.31 & 6.753 & 0.866 & 0.522 & 1535.46\\ 
\hline
\end{tabular}
\end{table}

\begin{table}
\caption{\label{tab:Table2}Potential parameters for 
$^{16}$O+$^{64}$Ni}
\begin{tabular}{cccccccc}\hline
E(MeV) & $N_R$ & WS & $R_w$ & $a_w$ & NORM &
$\chi^2$/point & $\sigma_R$(mb)\\
\hline
44.0 & 1.41 & 1452.50 & 6.846 & 0.434 & 0.976 & 1.248 & 500.95\\
     & 1.37 & 1150.60 & 6.846 & 0.449 & 0.976 & 1.418 & 504.03\\
     & 1.34 & 990.78 & 6.846 & 0.459 & 0.976 & 1.564 & 505.95\\
     & 1.27 & 694.80 & 6.846 & 0.484 & 0.976 & 2.031 & 510.40\\
     & 1.20 & 501.50 & 6.846 & 0.509 & 0.976 & 2.625 & 515.04\\
     & 1.14 & 370.60 & 6.846 & 0.534 & 0.976 & 3.329 & 519.43\\
     & 1.08 & 280.27 & 6.846 & 0.559 & 0.976 & 4.133 & 523.67\\
60.0 & 1.26 & 597.10 & 6.846 & 0.480 & 1.001 & 4.522 & 1208.94\\
     & 1.22 & 433.30 & 6.846 & 0.505 & 1.001 & 3.424 & 1215.79\\
     & 1.18 & 323.30 & 6.846 & 0.530 & 1.001 & 2.756 & 1222.59\\
     & 1.15 & 246.40 & 6.846 & 0.555 & 1.001 & 2.531 & 1229.16\\
     & 1.11 & 192.35 & 6.846 & 0.580 & 1.001 & 2.769 & 1235.80\\
     & 1.08 & 152.33 & 6.846 & 0.605 & 1.001 & 3.490 & 1242.00\\
     & 1.05 & 122.99 & 6.846 & 0.630 & 1.001 & 4.706 & 1248.57\\
\hline
\end{tabular}
\end{table}

 It is to be noted that with the chosen model potential our search procedure will only provide the crossing point for the imaginary
potential. No crossing will be observed in the real potentials as the
shapes and fall-off of these potentials are pre-fixed. Therefore, the 
crossing point of the imaginary potentials will be treated as the 
radius of potential sensitivity. It is evident from the figures that the
imaginary crossings are quite distinct and unambiguous for these systems
at all the energies studied.
As the imaginary potential through
dispersion relation generates the real polarization potential, it is 
justified to use the imaginary crossing radius in the study of radius
of potential sensitivity.

The observed phenomenon is that at near barrier
energies, the tightly bound projectiles like $^{16}$O, in principle, probe a unique radius
of the potential determined by the crossing point radius, and it is very close to the strong absorption radius. 
The variation of crossing
point radius or sensitive radius with incident energy is not significant. Therefore the evaluation
of the dispersion integral at the strong absorption radius is quite justified.
For the weekly bound projectiles, the behavior is
different. The crossings for $^{6}$Li+$^{58,64}$Ni are located in the vicinity of the strong
absorption radius for higher bombarding energies, 
but the values are larger by $\sim$20\% than the strong absorption
radius at lower bombarding energy. Similar observation for light targets has also been reported by Roubos, et al.,[4]. 
Hence care should be taken while evaluating the
dispersion relation in the investigation of energy dependence of effective
potential for loosely bound projectiles.
The crossing radii obtained for 
$^6$Li+$^{64}$Ni and $^{16}$O+$^{64}$Ni are compared in TableIII.

\begin{table}
\caption{\label{tab:table3}Crossing radii with energy for
$^6$Li+$^{64}$Ni and $^{16}$O+$^{64}$Ni}
\begin{tabular}{cccc}\hline
E(MeV) & Radius(fm) for & E(MeV) & Radius(fm) for\\
       & $^6$Li+$^{64}$Ni        &        & $^{16}$O+$^{64}$Ni\\ 
\hline
13.0 & 10.80 & 44.0 & 10.10\\
14.0 & 10.25 & 60.0 & 10.05\\
17.0 & 9.60 & &\\
19.0 & 8.90 & &\\
26.0 & 8.43\\\hline
\end{tabular}
\end{table}

An interesting aspect of the work in Ref.\cite{rou} is the observation of two crossing points at below or top of the barrier energies for $^{6}$Li+$^{208}$Pb and $^{16}$O+$^{208}$Pb systems.  
The authors have shown that the one at higher radius value corresponds to nearside scattering and the other at lower radius value corresponds to farside scattering. To identify the crossings associated with nearside and farside scattering we followed the prescriptions of Ref.\cite{rou}. We have performed our analysis in two steps for all the said four systems at top of the barrier energies. First we have fitted our elastic angular distributions considering the forward angle data only that is, $15^{o} \leq \theta_{c.m.} \leq 125^{o}$
, angles up to the point where the ratio of the angular distribution to Rutherford drops to $\sim$0.5. Next we have taken into account only the backward angle data, more specifically, $123^{o} \leq \theta_{c.m.} \leq 176^{o}$, to obtain the fit. We did not observe the "two crossings" situation for both $^{6}$Li+$^{58,64}$Ni and $^{16}$O+$^{58,64}$Ni systems at near barrier energies. In Fig.5 crossings associated with two different angular regions of the angular distributions for $^{6}$Li+$^{58}$Ni and $^{16}$O+$^{58}$Ni at near Coulomb barrier energies have been compared. The observed crossing for forward and backward angle data differ slightly but not enough to identify as two distinct crossings. Possibly the said decoupling between the nearside and farside scattering did not occur at this mass region.  

\begin{figure}
\vskip -1cm
\resizebox{\columnwidth}{!}{\includegraphics{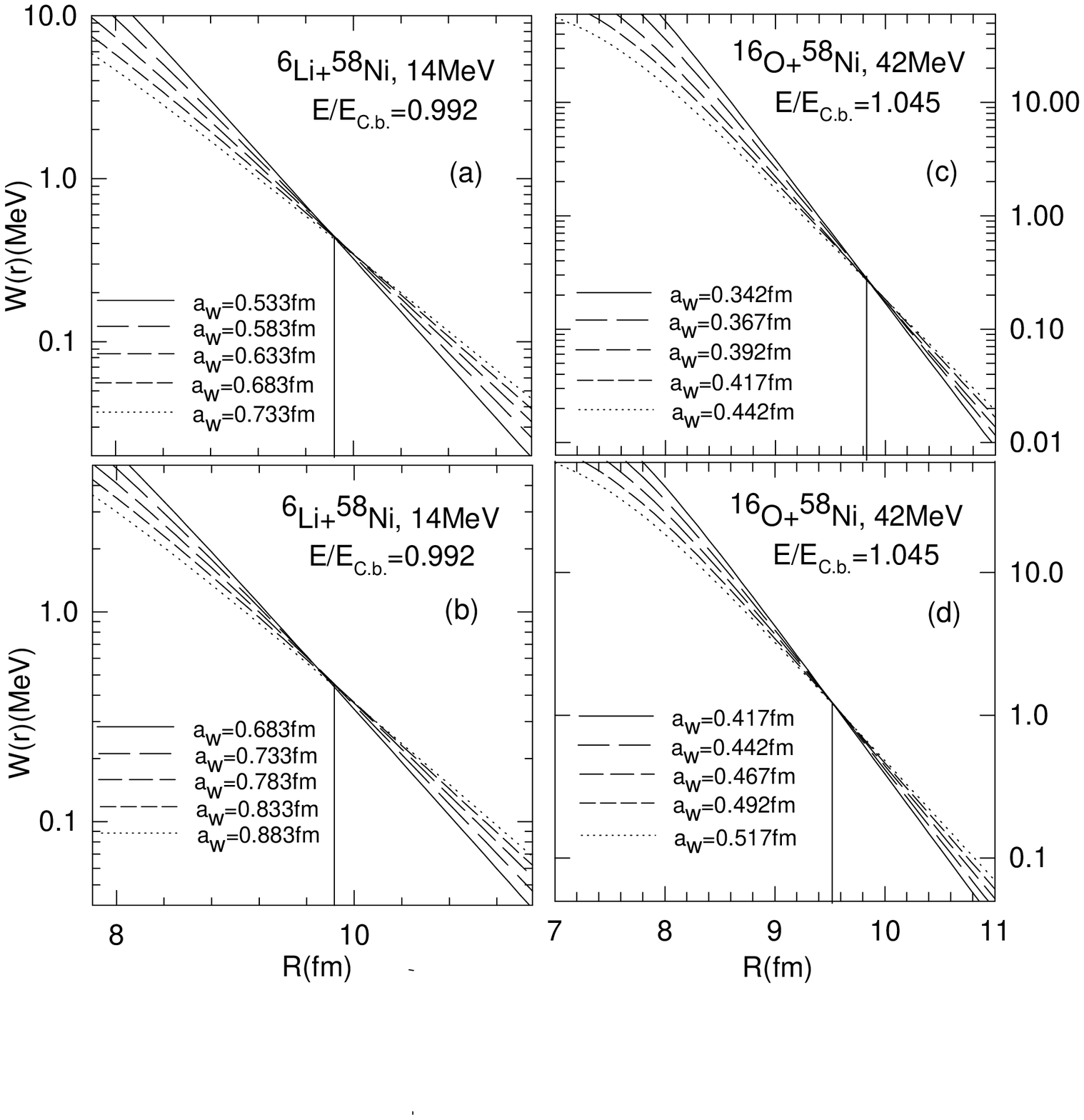}}
\vskip -3.2cm
\caption{\label{fig5}Crossing points obtained from fits to the forward angle data $(15^{o} \leq \theta_{c.m.} \leq 125^{o})$
 [(a)and(c)] and backward angle data $(123^{o} \leq \theta_{c.m.} \leq 176^{o})$ [(b)and(d)] of the elastic angular 
distributions of $^{6}$Li+$^{58}$Ni and $^{16}$O+$^{58}$Ni}
\end{figure}

The energy dependent nature of the crossing point radius for weakly bound $^{6}$Li+$^{58,64}$Ni systems has been depicted in Fig.3 where the crossing radii for $^{6}$Li+$^{64}$Ni at 14, 19 and 26MeV have been shown. As the energy goes higher the 
radius becomes smaller with enhanced absorption strength. The same energy dependence has been observed in case of $^{6}$Li+$^{58}$Ni system too. Interestingly these crossing radii are closer to the interaction distances at which the ratio $\sigma/\sigma_{Ruth}$ for those energies drops to 98\%. This possibly indicates that unlike the strongly bound projectiles where fusion at relatively lower radius dominates the absorption process at near barrier energies, the absorption for loosely bound projectile like $^{6}$Li is dominated by reactions at large separation. Breakup at large separation or single neutron transfer leading to unbound ejectiles could be possible reaction processes controlling the absorption on approaching the barrier.

In summary, we have performed a systematic radial sensitivity analysis of $^{16}$O+$^{58,64}$Ni and $^{6}$Li+$^{58,64}$Ni 
elastic scattering data. Two-crossing effect at the barrier has not been observed for any of the four systems studied. 
However, as pointed out by Roubos, et al.[4], to probe the existence of two crossings requires more experiments emphasizing 
the backward angle data with good statistics in the light mass targets. \\

\begin{acknowledgments}
The author would like to thank Prof. Subinit Roy for his generous
help and fruitful discussions to carry out this work. In addition,
the author gratefully acknowledges N. Keeley for providing with the
$^{16}$O+$^{58,64}$Ni data in tabular form. 
\end{acknowledgments}

\end{document}